\documentclass[journal=jacsat,manuscript=article]{achemso}

\usepackage{chemformula} 
\usepackage[T1]{fontenc} 
\usepackage[per-mode=symbol]{siunitx}
\usepackage{xcolor}
\usepackage[nolist]{acronym}
\usepackage{subcaption}
\usepackage{graphicx}
\usepackage{graphbox} 

\newcommand*\polyaffi{Department of Engineering Physics, École Polytechnique de Montréal, C.P. 6079, Succ. Centre-Ville, Montréal, Québec, Canada H3C 3A7}
\newcommand{\bostondept}{Department of Engineering, University of Massachusetts Boston, Boston, MA 02125, USA}
\newcommand{\pin}{$p$-$i$-$n$}
\newcommand{\gesn}[2 ]{Ge$_{#1}$Sn$_{#2}$}
\newcommand{\comments}[1]{}
\newcommand{\refc}[1]{\textcolor{blue}{#1}}

\author{Gérard Daligou}
\affiliation{\polyaffi{}}
\author{Mahmoud R. M. Atalla}
\affiliation{\polyaffi{}}
\author{Cédric Lemieux-Leduc}
\affiliation{\polyaffi{}}
\author{Anthony Nomezine}
\affiliation{\polyaffi{}}
\author{Simone Assali}
\affiliation{\polyaffi{}}
\author{Richard Soref}
\affiliation{\bostondept{}}
\author{Oussama Moutanabbir}
\email{oussama.moutanabbir@polymtl.ca}
\affiliation{\polyaffi{}}

\title{Mid-Infrared Thermal Radiation Harvesting using Uncooled Narrow Bandgap GeSn Thermophotovoltaic cell}

\abbreviations{IR,NMR,UV}
\keywords{American Chemical Society, \LaTeX}

\begin{document}

\acrodefplural{PCE}[PCEs]{power conversion efficiencies}
\begin{acronym}
  \acro{HS}{heterostructure}
  \acro{PV}{photovoltaic}
  \acro{PL}{photoluminescence}
  \acro{TPV}{thermophotovoltaic}
  \acro{SRH}{Shockley-Read-Hall}
  \acro{IQE}{internal quantum efficiency}
  \acro{EQE}{external quantum efficiency}
  \acro{PBA}{parabolic band approximation}
  \acro{TMM}{transfer matrix method}
  \acro{GTMM}{generalized transfer matrix method}
  \acro{EHP}{electron-hole pair}
  \acro{ETM}{equispaced thickness method}
  \acro{RPM}{random phase method}
  \acro{FF}{fill factor}
  \acro{PCE}{power conversion efficiency}
  \acro{ICPV}{interband cascade photovoltaic cells}
  \acro{ICL}{interband cascade laser}
  \acro{RP}{reduced-pressure}
  \acro{CVD}{chemical vapor deposition}
  \acro{VS}{virtual substrate}
  \acro{RIE}{reactive ion etching}
  \acro{ICP}{inductively coupled plasma}
  \acro{IPA}{isopropyl alcohol}
  \acro{BOE}{buffered oxide etch}
  \acro{PECVD}{plasma-enhanced chemical vapor deposition}
  \acro{RPCVD}{reduced pressure chemical vapor deposition}
  \acro{VF}{view factor}
  \acro{SLS}{stabilized light source}
  \acro{FTIR}{Fourier transform infrared}
  \acro{TEG}{thermoelectric generator}
  \acro{TEC}{thermal electrochemical cell}
  \acro{ICPV}{interband cascade photovoltaic}
  \acro{BSR}{back-surface reflector}
  \acro{IR}{infrared}
  \acro{XRD}{X-ray diffraction}
  \acro{RSM}{reciprocal space mapping}
  \acro{TEM}{transmission electron micrograph}
  \acro{APT}{atom probe tomography}
  \acro{EFA}{envelope function approximation}
  \acro{GeSn}{Germanium Tin}
  \acro{EDX}{Energy-dispersive X-ray}
  \acro{RT}{room temperature}
  \acro{SMU}{source measure unit}
  \acro{MWIR}{midwave-infrared}
\end{acronym}







\begin{abstract}
\Ac{TPV} cells are increasingly attractive for applications in industrial waste heat harvesting, aerospace energy management, and compact power generation. Deploying \ac{MWIR} \ac{TPV} in practical applications requires narrow-bandgap semiconductors that not only absorb low-energy photons but also integrate with scalable, low-cost platforms. Although high-performance \ac{TPV} devices have been demonstrated using III-V materials such as InAs, GaSb, and InGaAs(P), their use remains limited by cost and substrate size. With this perspective, narrow bandgap GeSn alloys are a promising alternative that extend group-IV absorption into the \ac{MWIR} while being silicon-compatible. Although the potential of GeSn \ac{TPV} cells  has been predicted, no experimental demonstration has been reported. Here, proof-of-concept \gesn{0.91}{0.09} \pin{} \ac{TPV} diodes (\qty{1}{\milli\meter} diameter) grown on silicon were fabricated and their performance was benchmarked against commercial InAs and extended-InGaAs devices. Measurements at \qty{300}{\kelvin} under \qty{2.3}{\micro\meter} laser and \qty{\sim 1500}{\kelvin} SiC Globar illumination revealed peak responsivity of \qty{\sim 0.2}{\ampere\per\watt} and an output power of \qty{\sim 0.41}{\milli\watt\per\centi\meter\squared}. These devices show trends comparable to those of the InAs diode under identical conditions, although at reduced absolute levels. To assess the intrinsic performance potential, Poisson-drift-diffusion modeling incorporating experimentally calibrated emitter emissivity predicts power densities exceeding \qty{1}{\watt\per\centi\meter\squared} under moderate \ac{MWIR} thermal illumination, indicating that the present devices operate far below their fundamental limits and are primarily constrained by defect-assisted recombination and transport losses.
These results establish GeSn as a scalable, silicon-compatible MWIR TPV platform and highlight a larger performance potential achievable through material and device optimization.
\end{abstract}

\section{Introduction}
The overlap of blackbody emission from sources in the \num{500}-\qty{2000}{\kelvin} range with the mid-infrared portion of the electromagnetic spectrum has spurred strong interest in narrow bandgap \acf{TPV} technologies\cite{meyer2025midinfraredsemiconductorphotonics}. By directly converting thermal radiation into electricity, \ac{TPV} devices are attractive for applications such as industrial waste heat harvesting, aerospace energy management, remote power generation, and portable devices \cite{luInAsThermophotovoltaicCells2018,chanPortableMesoscaleThermophotovoltaic2018,wangRadioisotopeThermophotovoltaicGenerator2020, rizzoComparisonTerahertzMicrowave2020a, yangNarrowBandgapPhotovoltaic2022,lapotinThermophotovoltaicEfficiency402022,chenReviewCurrentDevelopment2024a}. In this spectral range, narrow bandgap semiconductors are required to achieve efficient photon absorption, raising additional challenges in materials development, radiation-matter interaction, carrier transport, and device fabrication \cite{daligou2023group}.

\par The development of \ac{TPV} cells has historically mirrored advances in solar \ac{PV}, with early devices adapting conventional $p$-$n$ junction architectures for infrared absorption \cite{swansonSiliconPhotovoltaicCells1978,wedlockThermophotovoltaicEnergyConversion1963}. Yet, the distinct requirements of \ac{MWIR} \ac{TPV}, including operation with lower-energy photons, management of thermalization losses, and the need for robust spectral control, have driven the exploration of specialized emitter, filter, and device architectures \cite{chenReviewCurrentDevelopment2024a,roy-layindeHighefficiencyAirbridgeThermophotovoltaic2024}. In particular, transmissive spectral control using highly reflective back-surface mirrors has emerged as the most effective strategy to recycle sub-bandgap photons and enhance efficiency in far-field \ac{TPV} operation \cite{tervoEfficientScalableGaInAs2022b,chenReviewCurrentDevelopment2024a}.

\par At the material level, most high-efficiency TPV devices demonstrated to date have relied on III-V compounds such as GaSb\cite{fraasTPVGENERATORSUSING}, InGaAs(P)\cite{fanNearperfectPhotonUtilization2020,tervoEfficientScalableGaInAs2022b,roy-layindeHighefficiencyAirbridgeThermophotovoltaic2024}, and InAs\cite{luInAsThermophotovoltaicCells2018,selvidgeLargeAreaNearField2025}, owing to their favorable bandgaps and mature processing techniques. However, these materials face inherent limitations, including substrate high cost and small size, which restrict scalability for large-area deployment\cite{lemireGermaniumTinDiodeThermophotovoltaic2023}. These challenges have motivated growing interest in group-IV alternatives, particularly germanium-tin (GeSn) alloys\cite{daligou2023group}. By tuning the Sn content and strain state, GeSn offers access to direct bandgaps in the \ac{MWIR} while retaining compatibility with established silicon and germanium platforms\cite{moutanabbirMonolithicInfraredSilicon2021b}. This combination provides pathways toward cost-effective manufacturing, large wafer integration, and reduced material toxicity.

\par Recent progress in GeSn epitaxy and device processing has already enabled a variety of optoelectronic devices in the short- to the mid-wave infrared range \cite{atallaAllGroupIV2021b, atallaHighBandwidthExtendedSWIRGeSn2022a, bucaRoomTemperatureLasing2022, changMidinfraredResonantLight2022, chretienGeSnLasersCovering2019, chretienRoomTemperatureOptically2022, elbazUltralowthresholdContinuouswavePulsed2020b, joo1DPhotonicCrystal2021, jungOpticallyPumpedLowthreshold2022, li30GHzGeSn2021, liuSnContentGradient2022, luoExtendedSWIRPhotodetectionAllGroup2022b, marzbanStrainEngineeredElectrically2022, talamassimolaCMOSCompatibleBiasTunableDualBand2021, tranSiBasedGeSnPhotodetectors2019, xuHighspeedPhotoDetection2019, zhouElectricallyInjectedGeSn2020a}, underscoring the potential of the material for \ac{TPV} applications. Several theoretical and numerical studies have predicted the performance of GeSn-based \ac{TPV} devices, exploring bandgap tuning, layer thickness, and expected efficiencies under \ac{MWIR} illumination\cite{Conley2012,zhuGeSn0524EV2022,lemireGermaniumTinDiodeThermophotovoltaic2023,daligou2023group,tangComparisonOptimizedGeSn2023a}. However, to the best of our knowledge, no experimental demonstration of a GeSn \ac{TPV} device has been reported. Building on this motivation, the present work reports a proof-of-concept GeSn \ac{TPV} diode, demonstrating the feasibility of experimental radiation energy conversion in the \ac{MWIR} and providing the basis for future performance optimization.

\section{Results and discussion}

\subsection{Growth of GeSn epilayers}
GeSn \pin{} \acp{HS} were grown on 4-inch Si(100) substrates using \ac{RPCVD}, following similar growth sequence and parameters to those used in Ref.~\cite{atallaContinuousWaveGeSnLightEmitting2024}. Substrate preparation involved HF-based surface cleaning, followed by the growth of Ge \ac{VS}. The latter is obtained through a three-step process: growth of an initial 600-\qty{700}{\nano\meter}-thick Ge layer, followed by thermal cycling annealing up to \qty{800}{\celsius}, and a subsequent growth of an additional Ge layer to improve crystalline quality\cite{assaliAtomicallyUniformSnrich2018a}. The active region comprises a GeSn-based vertical \pin{} junction formed atop a four-layer graded GeSn buffer. The $p$-type \gesn{0.94}{0.06}, intrinsic \gesn{0.91}{0.09}, and $n$-type \gesn{0.95}{0.05} layers were deposited at \qty{335}{\celsius}, \num{305}-\qty{315}{\celsius}, and \qty{345}{\celsius}, respectively, using GeH$_4$ and SnCl$_4$ precursors. These growth temperatures were optimized to achieve the targeted Sn content and strain conditions required for a double \ac{HS} with a direct bandgap $i$-layer, despite potential fluctuations in Sn incorporation caused by partial strain relaxation and/or the presence of dopants \cite{assaliEnhancedSnIncorporation2019a,atallaHighBandwidthExtendedSWIRGeSn2022a,atallaContinuousWaveGeSnLightEmitting2024}. Doping was obtained using diborane (B$_2$H$_6$) and arsine (AsH$_3$) precursors, yielding active carrier concentrations above \qty{e19}{\centi\meter^{-3}}, as confirmed by capacitance-voltage measurements\cite{atallaContinuousWaveGeSnLightEmitting2024}. Two thin heavily doped $n$-type GeSn contact layers were deposited atop the \pin{} stack to reduce contact resistance and improve device performance. Note that the strain and microstructure of the as-grown double \ac{HS} are identical to those reported in Ref.~\cite{atallaContinuousWaveGeSnLightEmitting2024}, with layer thicknesses of \qty{346}{\nano\meter}, \qty{432}{\nano\meter}, and \qty{242}{\nano\meter}, for the $p$-, $i$-, and $n$-layers, respectively. The Ge-\ac{VS} exhibits an in-plane thermal mismatch-induced tensile strain of \qty{\sim 0.2}{\%}, while the GeSn buffer layers show relatively low compressive strain, reaching  about -0.2\% in the upper layers. Additionally, the $p$-layer is fully relaxed, the $i$-layer has a compressive strain of \qty{-0.24}\%, and the $n$-layer is slightly tensile strained, with a value of  \qty{\sim 0.3}\%, as measured by X-ray diffraction reciprocal space mapping. Besides, transmission electron microscopy (not shown) indicates high crystalline quality in the $i$- and $n$-layers, while the $p$-layer and underlying buffer contain dislocations, particularly near the Si/Ge-\ac{VS} interface.

\comments{Further structural and morphological details, including cross-sectional \ac{TEM} and \ac{XRD} analysis, are reported in Ref.~\cite{atallaContinuousWaveGeSnLightEmitting2024}.}

\subsection{GeSn \texorpdfstring{\pin{}}{p-i-n} devices}
\begin{figure}[htb]
  \centering
  \includegraphics[width=0.9\textwidth]{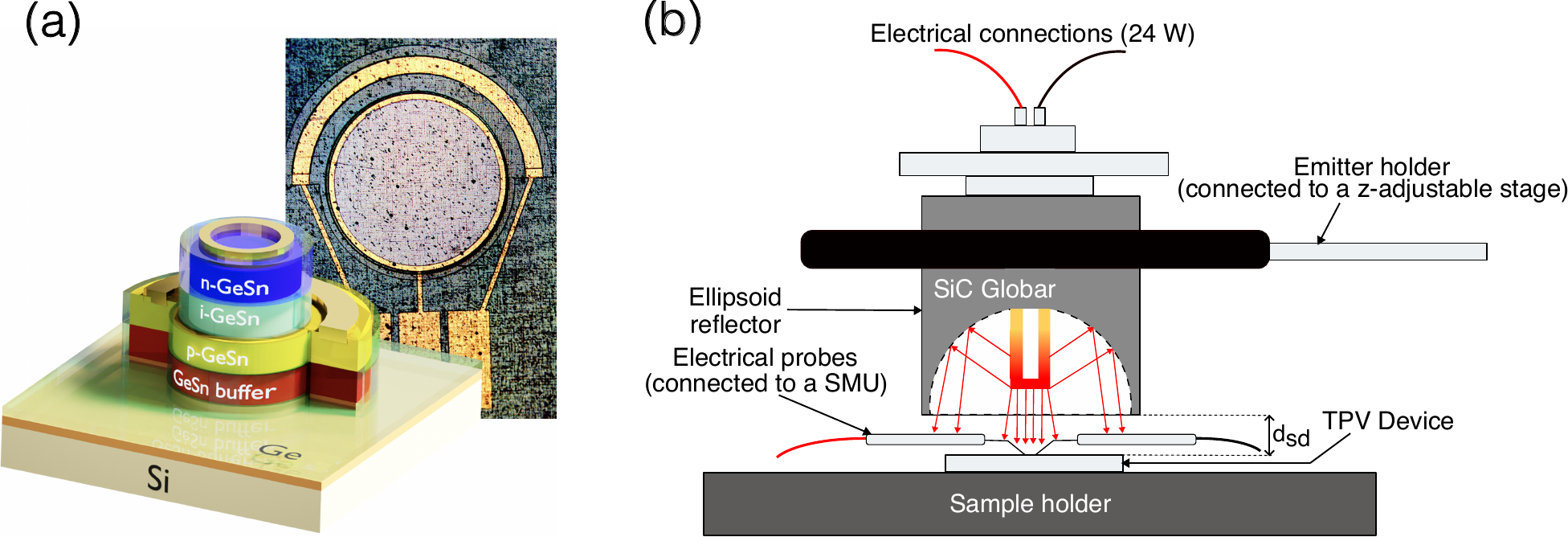}
  \caption{\gesn{0.94}{0.06}/\gesn{0.91}{0.09}/\gesn{0.95}{0.05} \pin{} device structure and SiC-based illumination setup. (a) Schematic and optical micrograph of a representative \pin{} \ac{TPV} device. The glossy-like material covering the device represents the SiO$_2$ passivation layer, whereas the gold-like structure on top on both the $p$- and $n$- GeSn layers are the metallic contacts. (b) Schematic of the measurement configuration for SiC-based \ac{TPV} illumination, including beam delivery and device mounting stages.}
  \label{fig:devSetupProps}
\end{figure}

\comments{This design was motivated by the need to extend previously reported vertical GeSn photodiodes to larger-scale devices suited for \ac{TPV} characterization\cite{atallaHighBandwidthExtendedSWIRGeSn2022a}.}
The GeSn photodiodes studied in this work were fabricated using a circular mesa vertical \pin{} architecture, with an active area diameter of up to \qty{1}{\milli\meter}, as shown in Figure \ref{fig:devSetupProps}(a). While state-of-the-art \ac{TPV} cells typically employ front metal gridlines combined with a full-area back contact to minimize series resistance \cite{chenReviewCurrentDevelopment2024a,burgerPresentEfficienciesFuture2020, roy-layindeHighefficiencyAirbridgeThermophotovoltaic2024}, the present design was intentionally chosen to minimize fabrication-related uncertainties and enables a focused assessment of the intrinsic junction behavior under \ac{TPV}-relevant illumination conditions. The circular mesa geometry further provides a well-defined active area and uniform illumination, facilitating reproducible electrical and optical analyses. Although not optimized for maximum \ac{TPV} efficiency, this architecture establishes a robust baseline for evaluating the potential of GeSn \pin{} devices for \ac{MWIR} \ac{TPV} applications.
The process began with GeSn \pin{} surface cleaning using HF/HCl/H$_2$O (1:1:2) solution. Photolithography was then performed to define the top mesa features, followed by anisotropic etching of the top $n$-type GeSn layer down to the $p$-type region using an \ac{ICP}-\ac{RIE} system with a Cl$_2$/N$_2$/O$_2$ chemistry. A second photolithography and etch step was carried out to further define an underlying mesa down to the Ge virtual substrate, effectively creating a double-mesa structure that enhances device isolation and reduces parasitic contact effects. After each dry etching step, sidewall passivation was performed using a brief immersion in HCl/HF (1:1) to remove surface oxides and suppress leakage paths. A \qty{\sim 1.5}{\micro\meter} thick SiO$_2$ passivation layer was then deposited by \ac{PECVD} to provide insulation and support low-capacitance contacts. Contact vias were patterned and opened through buffered oxide etching. Finally, metal contacts were formed by e-beam evaporation of a Ti/Au/Ag/Au multilayer stack, where Ag served to fill SiO$_2$ topography and minimize the total thickness of gold required for reliable contact pads. \comments{This process builds upon fabrication strategies reported in prior GeSn \pin{} devices by Atalla \latin{et al}.~\cite{atallaHighBandwidthExtendedSWIRGeSn2022a}}

\par
\begin{figure}[htb]
  \centering
  \includegraphics[width=0.9\textwidth]{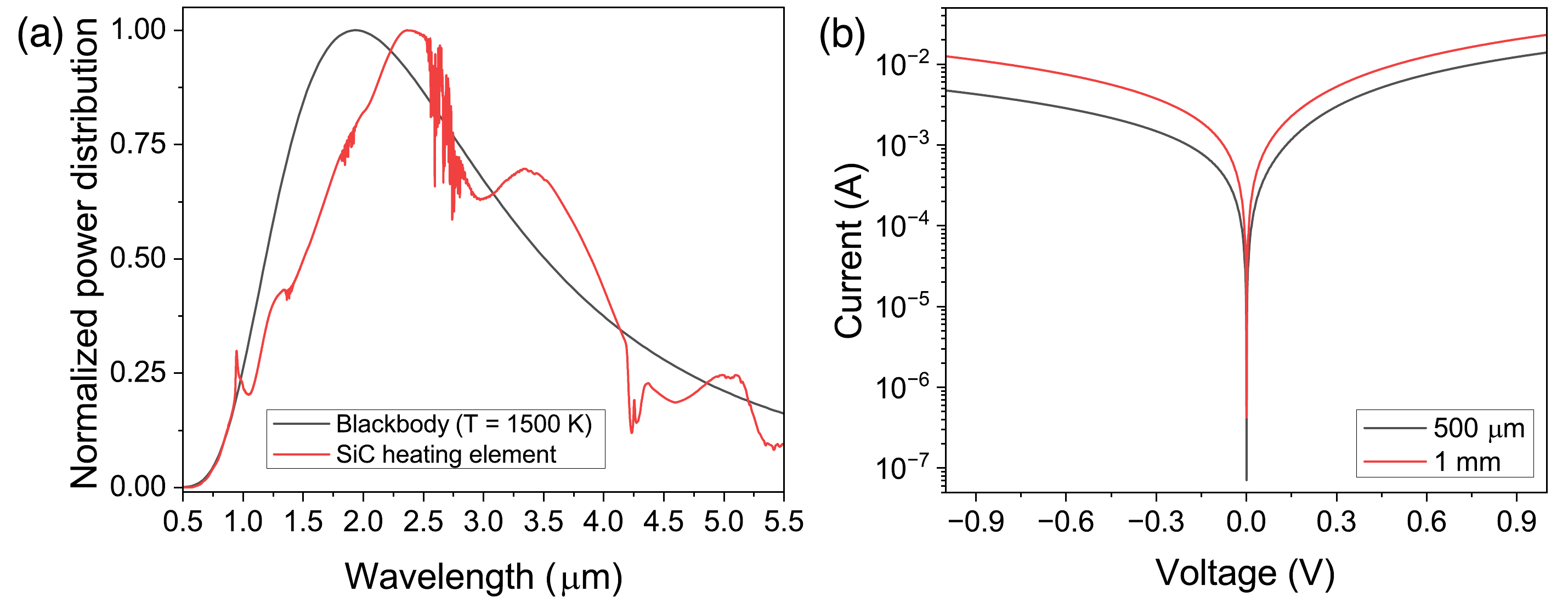}
  \caption{Broadband source characterization and dark I-V device characteristics. (a) Emission spectrum of the SiC heating element at \qty{\sim 1500}{\kelvin}, highlighting its broadband gray-body distribution extending across the mid-infrared \cite{thorlabsSiC}. (b) I-V characteristics under dark condition for devices with \qty{500}{\micro\meter} and \qty{1}{\milli\meter} active area diameter.}
  \label{fig:baselineGraphs}
\end{figure}

 Current-voltage (I-V) measurements were performed at \qty{300}{\kelvin} using a \ac{SMU} connected to a probe station. The fabricated GeSn devices were evaluated under two illumination regimes: (i) laser-based power beaming and (ii) broadband \ac{TPV} operation. For power beaming, a \qty{2.3}{\micro\meter} laser was directed onto the device after passing through optical components, including beam splitters, mirrors, and focusing lenses. For broadband operation, a SiC heating element served as the infrared emitter. This emitter is housed in an ellipsoid reflector to increase the optical output, and help in defining the beam size. The device was mounted on a sample holder equipped with $x$-$y$ linear translation for alignment, while the emitter was positioned on a $z$-adjustable stage to vary the source-detector distance ($d_{sd}$). This geometry enables tuning of the emitter-detector \ac{VF}, defined as the overlap ratio between the source and detector active areas, which directly influences the optical coupling efficiency\cite{limLimitsEnergyConversionEfficiency2023b}. Figure \ref{fig:devSetupProps}(b) presents a schematic of the experimental setup. The SiC emitter, with a nominal color temperature of \qty{\sim 1500}{\kelvin}, provides a gray-body spectrum extending from 0.5 to \qty{9.0}{\micro\meter} based on the data provided by the supplier.\cite{thorlabsSiC}.  The spectrum of the SiC heating element was measured from 0.4 to \qty{5.6}{\micro\meter}, as highlighted in Figure \ref{fig:baselineGraphs}(a). The structures in the measured light source spectra are due to absorption from various molecules such as H$_2$O and CO$_2$\cite{thorlabsSiC}. Although manufacturer specifications do not include the spectral emissivity $\varepsilon (\lambda)$, this parameter is essential for accurate estimation of the emitted power. In practice, $\varepsilon (\lambda)$ must be determined through calibration against blackbody standards under comparable conditions\cite{fanNearperfectPhotonUtilization2020}. 

\par \comments{The performance of the fabricated GeSn devices was assessed through spectral responsivity and current-voltage (I-V) measurements conducted at \ac{RT}.} 
As a baseline, dark I-V characteristics were measured for GeSn devices with active areas of \qty{500}{\micro\meter} and \qty{1}{\milli\meter}, as shown in Figure \ref{fig:baselineGraphs}(b). Both devices exhibited relatively low rectification ratios across the applied bias range. Specifically, the \qty{500}{\micro\meter} device achieved a rectification ratio of approximately 2.48 at \qty{0.5}{\volt}, whereas the \qty{1}{\milli\meter} device reached only \num{\sim 1.61} at the same voltage. These observations reflect the impact of growth-related defects, which are more pronounced in larger-area devices. Reducing device dimensions is expected to enhance rectification significantly. For instance, prior work on similar GeSn structures demonstrated that decreasing the diameter from \qty{120}{\micro\meter} to \qty{28}{\micro\meter} can increase the rectification ratio by one to two orders of magnitude\cite{atallaHighBandwidthExtendedSWIRGeSn2022a}. While smaller diameters (down to \qty{10}{\micro\meter}) enable high-speed operation at \ac{RT},\cite{Atalla2024} they are less favorable for power-generation applications, such as thermal radiation harvesting, where larger active areas are desirable. Thus, GeSn \ac{TPV} devices require a careful balance between material quality, heterostructure design, and device size to optimize performance for their intended application.  

\begin{figure}[htb]
  \centering
  \includegraphics[width=0.9\textwidth]{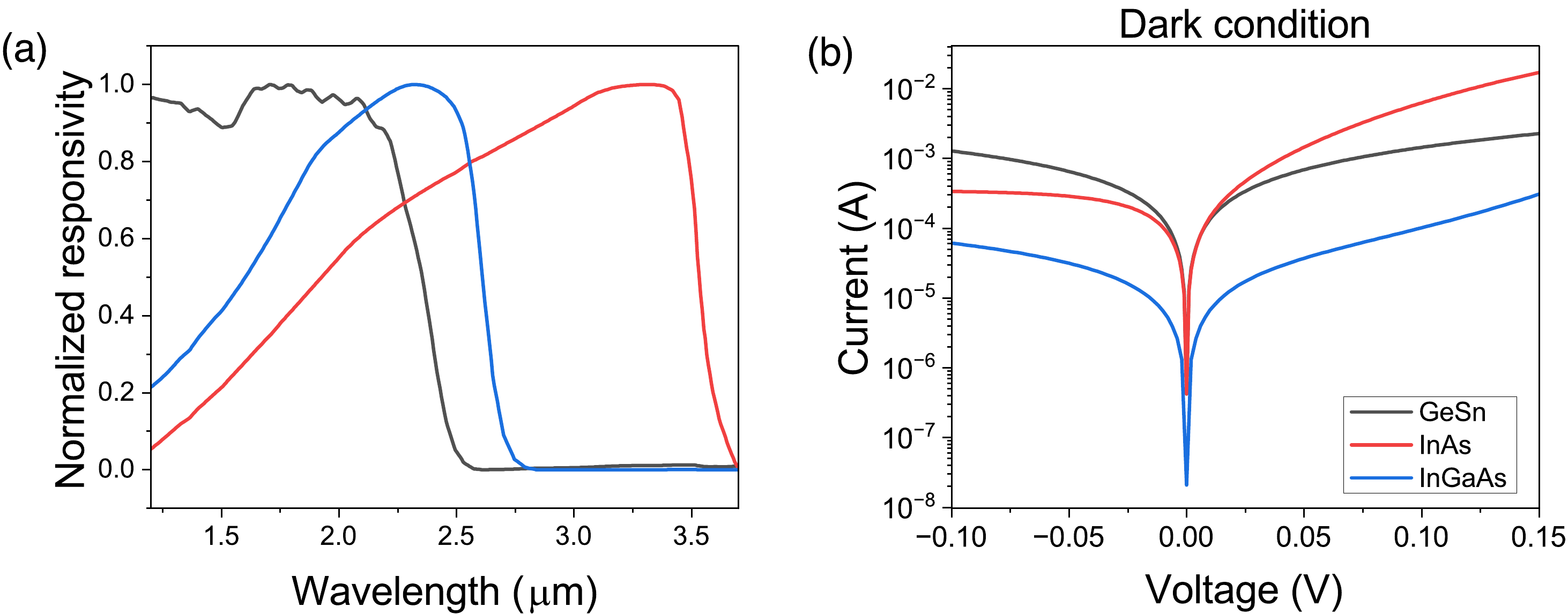}
  \caption{(a) Spectral responsivity at \qty{300}{\kelvin} of GeSn, InAs, and extended-InGaAs photodiodes with identical active area diameters (\qty{1}{\milli\meter}), measured, at \qty{0}{\volt}, using \ac{FTIR} spectroscopy and calibrated against InAs. The responsivities were normalized to highlight the trends and the cutoff wavelengths. (b) Dark current-voltage (I-V) characteristics of the same devices at \qty{300}{\kelvin}, showing comparable current levels for GeSn and InAs but lower rectification ratios compared to extended-InGaAs.}
  \label{fig:RspDark}
\end{figure}

\begin{figure}[htb]
  \centering
  \includegraphics[width=0.9\textwidth]{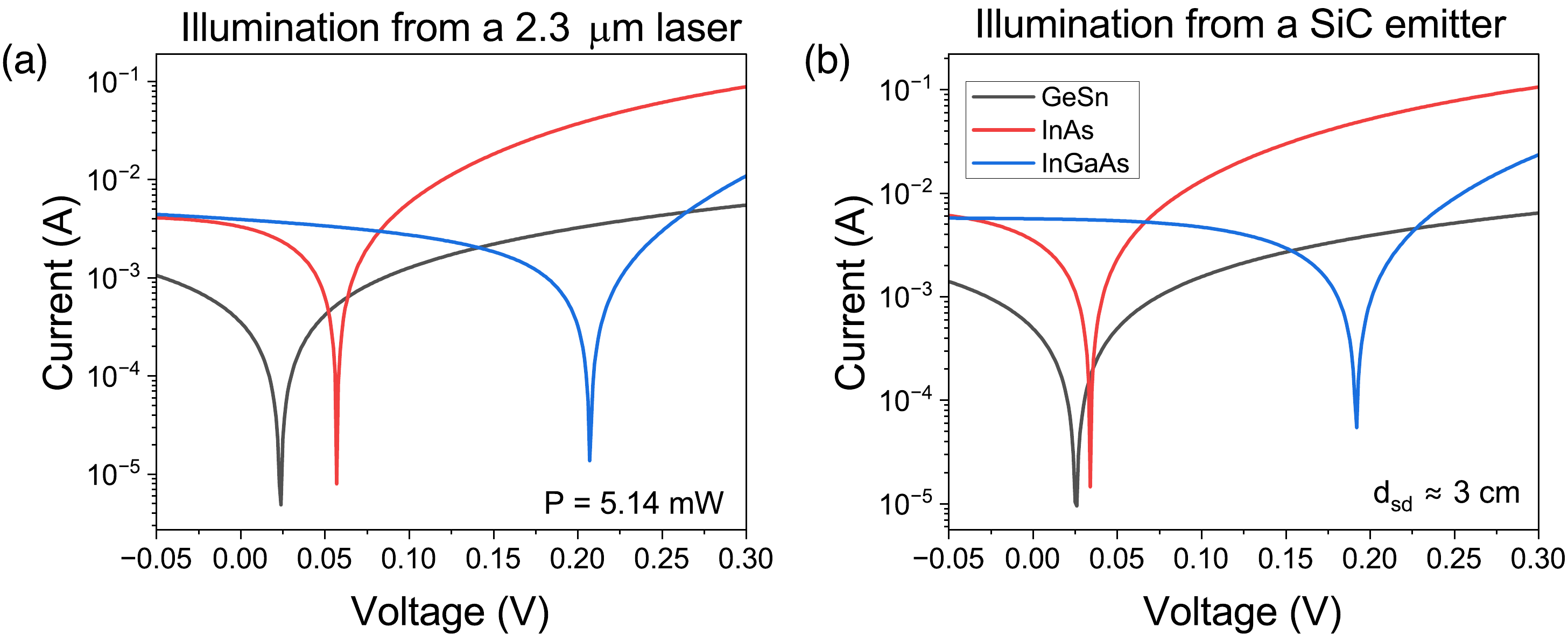}
  \caption{(a) I-V response of GeSn, InAs, and extended-InGaAs photodiodes under \qty{2.3}{\micro\meter} laser illumination, demonstrating photoresponse and output power density differences among the devices. (b) I-V characteristics under broadband mid-infrared radiation from a \qty{\sim 1500}{\kelvin} SiC emitter positioned \qty{3}{\centi\meter} above the detectors, reproducing trends observed under laser illumination.}
  \label{fig:IllumIV}
\end{figure}

With these configurations established, the GeSn diodes, with \qty{1}{\milli\meter} active area diameter, were measured alongside state-of-the-art commercial InAs and extended-InGaAs photodiodes of identical geometry, enabling direct comparison of performance under identical conditions.
I-V characteristics were obtained in the dark, under \qty{2.3}{\micro\meter} laser illumination ($P_{in}$ = \qty{5.14}{\milli\watt}), and under broadband infrared emission from a SiC Globar positioned \qty{\sim 3}{\centi\meter} above the devices. Relative spectral responsivities ($R_{sp}$) were measured, at \qty{0}{\volt}, using a \ac{FTIR} spectrometer and calibrated against the InAs photovoltaic detector, which has a well-characterized response. The results, summarized in Figures \ref{fig:RspDark} and \ref{fig:IllumIV}, highlight key differences in current levels, responsivity, and overall performance. The GeSn device exhibits a peak responsivity of \qty{\sim 0.2}{\ampere\per\watt} at \qty{\sim 1.7}{\micro\meter}, remaining relatively constant up to \qty{2.1}{\micro\meter} before gradually decreasing and cutting off near \qty{\sim 2.6}{\micro\meter}. By comparison, the InAs detector peaks at \qty{\sim 1}{\ampere\per\watt} at \qty{3.31}{\micro\meter} with a cutoff around \qty{3.65}{\micro\meter}, while the ext-InGaAs diode peaks at \qty{\sim 1.52}{\ampere\per\watt} at \qty{2.31}{\micro\meter} and cuts off near \qty{2.8}{\micro\meter}. Under dark conditions (Figure \ref{fig:RspDark}(b)), the GeSn device shows a dark current levels similar to those of the InAs detector at \qty{0}{\volt} (\qty{\sim 0.28}{\micro\ampere}), although with a lower rectification ratio. Both devices exhibit dark currents approximately one order of magnitude higher than that of the ext-InGaAs detector, which records \qty{\sim 34.5}{\nano\ampere} at \qty{0}{\volt} with a rectification ratio reaching \num{\sim 4e2} at \qty{0.5}{\volt}.

\par Under \qty{2.3}{\micro\meter} laser illumination, all devices displayed clear photocurrent generation, confirming their photo-response. The GeSn device produced a short-circuit current ($I_{sc}$) of approximately \qty{0.353}{\milli\ampere}, nearly three orders of magnitude higher than its dark current at \qty{0}{\volt}, with an open-circuit voltage ($V_{oc}$) of \qty{\sim 24}{\milli\volt}, corresponding to an output power density of \qty{0.27}{\milli\watt\per\centi\meter\squared}. The InAs detector showed higher $I_{sc}$ and $V_{oc}$ under identical conditions, resulting in an output power density of $\sim$\qty{7.9}{\milli\watt\per\centi\meter\squared}, while the ext-InGaAs detector delivered the highest output with $I_{sc}$\qty{\approx 3.97}{\milli\ampere}, $V_{oc}$\qty{\approx 0.21}{\volt}, and a peak electrical output power density of \qty{37}{\milli\watt\per\centi\meter\squared}.\comments{\refc{Nevertheless, this value remains relatively low for practical power generation, indicating that higher optical flux and improved thermal management would be necessary to achieve meaningful conversion efficiency for both GeSn and InAs}.} These differences are consistent with the higher responsivity of ext-InGaAs at the excitation wavelength. When illuminated with the broadband SiC emitter (Figure \ref{fig:devSetupProps}(b)), all devices showed enhanced photocurrents compared to dark conditions, following the same relative performance trends as under laser excitation. The GeSn detector achieved an output power density of \qty{\sim 0.41}{\milli\watt\per\centi\meter\squared}, while the InAs and ext-InGaAs detectors reached \qty{\sim 4.3}{\milli\watt\per\centi\meter\squared} and \qty{\sim 64}{\milli\watt\per\centi\meter\squared}, respectively. The observed differences in current levels and slope between laser and SiC illumination can be attributed to the distinct spectral distribution and optical flux of the two sources.

\par The observed performance differences among the three devices can be primarily attributed to material maturity, processing optimization, and inherent device properties. The GeSn diode, as a first proof-of-concept \ac{MWIR} \ac{TPV} device, exhibits measurable photocurrent and energy conversion, although dark currents and voltage output remain below those of the commercial InAs and ext-InGaAs detectors. The InAs device benefits from established growth and fabrication techniques, yielding improved photoresponse, while the ext-InGaAs detector achieves the highest performance under both illumination conditions, consistent with its optimized material quality and responsivity at the wavelengths tested.


\par To further quantify the intrinsic performance potential of GeSn-based \ac{TPV} devices and to contextualize the experimental results, numerical modeling was employed. While experimental measurements provide direct insight into device operation under thermal illumination, they do not isolate the relative contributions of material quality, recombination mechanisms, and carrier transport limitations. Device-level simulations can provide a complementary framework to establish realistic efficiency benchmarks and to evaluate the achievable power conversion efficiency of GeSn TPV cells under mid-infrared thermal radiation. The modeling approach adopted in this work is based on a one-dimensional coupled Poisson-drift-diffusion framework previously developed for \ac{TPV} devices\cite{daligou2023group}. The simulated \pin{} GeSn structure corresponds directly to the fabricated device, with layer thicknesses, doping concentrations, and material parameters selected to match the measured devices. The optical generation rate within the multilayer GeSn stack is calculated using the optical transfer matrix method, while the electrical model explicitly accounts for intrinsic and extrinsic recombination mechanisms relevant to narrow-bandgap group-IV semiconductors. Note that parasitic resistances are neglected and ideal ohmic contacts are assumed; therefore, the simulated performance represents an upper bound for the fabricated devices. The spectral emissivity of the SiC thermal emitter was taken from the work of Fan et al.\cite{fanNearperfectPhotonUtilization2020}, where the same lamp geometry and operating conditions were used and the emissivity was experimentally calibrated. Using these emissivity data, the incident photon flux and power density on the \ac{TPV} device were calculated and used as input to the model. For the experimental configuration, the incident power density was estimated to be approximately \qty{16}{\watt\per\centi\meter\squared}.

\par Using mobility, surface recombination rates, Auger coefficients, and the SRH capture lifetimes extracted from Ge as a practical baseline, due to the lack of detailed experimental parameters of GeSn materials\cite{caugheyCarrierMobilitiesSilicon1967,Liu2007,virgilio2013,bosiGermaniumEpitaxyIts2010}, the modeled GeSn device exhibits a short-circuit current of approximately \qty{ 34}{\milli\ampere} and an open-circuit voltage of approximately \qty{0.41}{\volt} , corresponding to a peak electrical output power density of \qty{\sim 1.37}{\watt\per\centi\meter\squared}. Despite the moderate incident power density, the simulated device performance significantly exceeds the experimentally measured values, with $I_{sc}$, $V_{oc}$, and $P_{max}$ being higher by factors of approximately 69, 16, and 3000, respectively. This large difference indicates that the fabricated GeSn \ac{TPV} devices operate far from their intrinsic performance limits and that non-ideal material and transport effects dominate the measured response. In particular, the simulations neglect parasitic resistances, non-uniform illumination, and contact losses, all of which are expected to reduce carrier collection and photovoltage in practical devices. More importantly, the deviation between modeled and measured characteristics points toward strong non-radiative recombination losses in the GeSn absorber. At the injection and doping levels relevant to the present \ac{TPV} conditions, Auger recombination is not expected to be the dominant loss mechanism. Instead, \ac{SRH} recombination is expected to play a central role in governing carrier lifetime, collection efficiency, and photovoltage. In GeSn alloys, \ac{SRH} recombination is further exacerbated by the presence of misfit dislocations, point defects, and interface states associated with strain relaxation and heteroepitaxial growth on Ge/Si substrates \cite{moutanabbirMonolithicInfraredSilicon2021b,Giunto2024_DefectsGeGeSn}.

\par To illustrate the impact of defect-assisted recombination on GeSn \ac{TPV} performance, simulations were performed for different \ac{SRH} capture lifetimes $\tau_{n_0}$, $\tau_{p_0}$. Because carrier-specific SRH parameters in GeSn are not well established, equal electron and hole capture lifetimes ($\tau_{n_0}$ = $\tau_{p_0}$ = $\tau_0$) are assumed, with $\tau_0$ treated as an effective bulk lifetime reflecting the defect density. In addition to the Ge-like baseline ($\tau_0 =$ \qty{2e-6}{\second}) mentioned above, two additional $\tau_0$ values have been chosen to represent moderate defect density ($\tau_0 =$ \qty{1e-8}{\second}) and defect-dominated regime ($\tau_0 \sim$ \qty{1e-10}{\second}). These values span the range expected for heteroepitaxial GeSn alloys and enable the estimation of the progressive degradation of $V_{oc}$, $I_{sc}$, and output power density with increasing non-radiative recombination\cite{moutanabbirMonolithicInfraredSilicon2021b,Julsgaard2020}. Reducing $\tau_0$ to \qty{1e-8}{\second} leads to a decrease in $V_{oc}$ to \qty{0.28}{\volt} and a corresponding reduction in peak power density to \qty{\sim 0.71}{\watt\per\centi\meter\squared}, while $I_{sc}$ decreases more moderately to \qty{ 29.4}{\milli\ampere} . In the defect-dominated regime, $V_{oc}$ collapses to \qty{\sim 0.09}{\volt} and $P_{max}$ is reduced by approximately 1.5 orders of magnitude compared to the Ge-like case, and $I_{sc}$ reached a value of \qty{10.14}{\milli\ampere}. These results highlight the dominant role of SRH recombination in limiting the voltage and power output of present GeSn TPV devices.

\par Finally, it is also important to mention that one of the primary limitations that need to be addressed in \ac{TPV} studies is rooted in the difficulty to accurately estimate the power emitted by the SiC source and, consequently, the \ac{PCE} of the devices. The lack of a standardized method for estimating \ac{PCE} makes comparisons across studies challenging and slows the systematic development of \ac{TPV} devices\cite{tervoEfficientScalableGaInAs2022b,lopezThermophotovoltaicConversionEfficiency2023,roy-layindeHighefficiencyAirbridgeThermophotovoltaic2024}. Two main strategies have been broadly adopted for estimating absorbed power in \ac{TPV} systems\cite{mahorterThermophotovoltaicSystemTesting2003,lopezThermophotovoltaicConversionEfficiency2023,narayanPlatformAccurateEfficiency2021}. The first relies on optical characterization, measuring the spectral emissivity of the emitter and the reflectance of the cell to estimate net absorbed radiation\cite{mahorterThermophotovoltaicSystemTesting2003}. This approach requires careful consideration of geometric effects, such as the emitter-cell view factor, and the angle- and temperature-dependent optical properties of both surfaces\cite{mahorterThermophotovoltaicSystemTesting2003,narayanPlatformAccurateEfficiency2021}. While relatively straightforward, this approach is highly sensitive to uncertainties in reflectance and assumes idealized radiative exchange conditions, limiting its accuracy for realistic device configurations\cite{tervoEfficientScalableGaInAs2022b,narayanPlatformAccurateEfficiency2021}. The second approach relies on calorimetry, quantifying absorbed energy by measuring heat flow through the device\cite{tervoEfficientScalableGaInAs2022b,lopezThermophotovoltaicConversionEfficiency2023,lapotinThermophotovoltaicEfficiency402022,swansonSiliconPhotovoltaicCells1978}. The principle is that all the absorbed incident energy is either dissipated as heat or converted to electrical power. This method is best suited for systems with a high emitter-cell view factor, where thermal losses can be monitored directly, but its practical implementation is challenging due to thermal management requirements and geometric constraints\cite{lopezThermophotovoltaicConversionEfficiency2023}. Both approaches highlight the importance of careful emitter-device alignment, thermal control, and accurate characterization of optical and thermal properties.




\section{Conclusion}

In this work, proof-of-concept GeSn-based \ac{TPV} diode is demonstrated, establishing GeSn as a viable silicon-compatible platform for \ac{MWIR} energy conversion. A \qty{1}{\milli\meter}-diameter vertical GeSn \pin{} device grown on silicon was fabricated and benchmarked against commercial InAs and extended-InGaAs detectors under both monochromatic (\qty{2.3}{\micro\meter}) and broadband thermal (\qty{\sim 1500}{\kelvin} SiC) illumination.  The measured peak responsivity and output power densities are \qty{\sim 0.2}{\ampere\per\watt} and \qty{\sim 0.41}{\watt\per\centi\meter\squared}, respectively.  GeSn devices were found to exhibit trends comparable to commercial InAs device under identical experimental conditions, although at reduced absolute levels.
GeSn TPV performance is limited primarily by defect-assisted recombination, non-ideal transport, elevated dark currents, and contact effects, highlighting the need for continued advances in epitaxial growth, heterostructure design, and contact engineering. To quantify the intrinsic performance potential, device-level Poisson-drift-diffusion simulations were performed.  The modeling predicts power densities exceeding \qty{1}{\watt\per\centi\meter\squared} under moderate exposure to mid-infrared thermal radiation, indicating that the current GeSn devices operate far from their optimal performance and that substantial gains are achievable through lifetime and defect-density improvements.
At the system level, emitter-detector coupling and standardized efficiency measurements remain essential to enable reliable benchmarking across \ac{TPV} platforms. Overall, these results provide the experimental validation of GeSn \ac{TPV} devices and reveal a large untapped performance margin. With continued progress in material quality, device architecture, and system integration, GeSn is poised to complement III-V TPV technologies and enable scalable, silicon-compatible \ac{MWIR} energy harvesting platforms.

\begin{acknowledgement}
The authors acknowledge support from NSERC Canada, Canada Research Chairs, Canada Foundation for Innovation, PRIMA Qu\'ebec, Defence Canada (Innovation for Defence Excellence and Security, IDEaS), the European Union’s Horizon Europe research and innovation program under Grant Agreement No 101070700 (MIRAQLS), and the Air Force Office of Scientific and Research Grant No. FA9550-23-1-0763.

\end{acknowledgement}

\bibliography{bibliography}

\end{document}